\documentclass[letterpaper,twocolumn,nofootinbib,aps,superscriptaddress]{revtex4-1}

\usepackage{amsmath}
\usepackage{amssymb}
\usepackage{hyphenat}
\usepackage{amsthm}
\usepackage{color}
\usepackage{bbm}
\usepackage{bm}
\usepackage{upgreek}
\usepackage{enumitem}
\usepackage{crossreftools}
\usepackage[hyperfootnotes=false]{hyperref}
\usepackage{footnotebackref}
\usepackage{graphicx}
\usepackage[caption=false]{subfig}
\usepackage[figure]{hypcap}
\usepackage{pdftexcmds}
\usepackage{mathrsfs}

\DeclareFontFamily{U}{cbgreek}{}
\DeclareFontShape{U}{cbgreek}{m}{n}{
        <-6>    grmn0500
        <6-7>   grmn0600
        <7-8>   grmn0700
        <8-9>   grmn0800
        <9-10>  grmn0900
        <10-12> grmn1000
        <12-17> grmn1200
        <17->   grmn1728
      }{}
\DeclareFontShape{U}{cbgreek}{bx}{n}{
        <-6>    grxn0500
        <6-7>   grxn0600
        <7-8>   grxn0700
        <8-9>   grxn0800
        <9-10>  grxn0900
        <10-12> grxn1000
        <12-17> grxn1200
        <17->   grxn1728
      }{}

\makeatletter
\newcommand{\normalorbold}{%
  \ifnum\pdf@strcmp{\math@version}{bold}=\z@ bx\else m\fi
}
\makeatother

\newtheorem{theorem}{Theorem}

\newtheorem{obsm}[theorem]{Observation}

\newtheorem{clm}[theorem]{Claim}
\newtheorem*{note}{Note}
\newtheorem*{axm}{Axiom}
\newtheorem{cls}{Clause}

\def\gbm#1{{\let\phi\upphi \let\lambda\uplambda \let\mu\upmu \let\rho\uprho \let\sigma\upsigma \let\tau\uptau \let\theta\uptheta \let\eta\upeta \bm{#1}}}

\newcommand*{\rA}{\mathrm{A}}

\newcommand*{\rD}{\mathrm{D}}

\newcommand*{\db}{\overline{\mathrm{D}}}
\newcommand*{\rE}{\mathrm{E}}
\newcommand*{\eb}{\overline{\mathrm{E}}}
\newcommand*{\rF}{\mathrm{F}}
\newcommand*{\rG}{\mathrm{G}}
\newcommand*{\fb}{\overline{\mathrm{F}}}
\newcommand*{\hb}{\overline{\mathrm{h}}}
\newcommand*{\rL}{\mathrm{L}}
\newcommand*{\lb}{\overline{\mathrm{L}}}
\newcommand*{\rM}{\mathrm{M}}
\newcommand*{\rR}{\mathrm{R}}
\newcommand*{\rS}{\mathrm{S}}
\newcommand*{\tb}{\overline{\mathrm{t}}}
\newcommand*{\rW}{\mathrm{W}}
\newcommand*{\wb}{\overline{\mathrm{W}}}

\newcommand*{\ket}[1]{\left|#1\right\rangle}
\newcommand*{\bra}[1]{\left\langle #1\right|}

\newcommand*{\fr}[2]{\frac{#1}{#2}}

\newcommand{\be}{\begin{equation}}
\newcommand{\ee}{\end{equation}}
\newcommand{\n}{\textendash}
\newcommand{\m}{\textemdash}

\hypersetup{
    unicode=true,          
    pdffitwindow=false,     
    pdfstartview={FitH},    
}

\newcommand{\frp}{\hyperlink{cite.FR18}{FR}}

\begin{document}
\title{Agents governed by quantum mechanics can use it intersubjectively and consistently}
\date{\today}

\author{Varun Narasimhachar}
\email{nvarun@ntu.edu.sg}
\affiliation{
School of Physical and Mathematical Sciences, Nanyang Technological University, 637371 Singapore, Singapore
}

\begin{abstract} 
Following Frauchiger and Renner's discovery of a conflict between quantum mechanics and certain commonsense reasoning axioms, much work has gone into finding alternative axiomatizations that can avoid the conflict. However, this body of work is largely based on specific interpretations of quantum mechanics, and at times employs specialized formalism that may be inaccessible to mainstream quantum information experts. Taking an interpretation\hyp agnostic approach, we propose a simple operational principle called \emph{superpositional solipsism} to aid rational agents in making situational inferences. We show that the principle leads to sound inferences in all operationally\hyp relevant instances. Along the way, we discuss certain subtleties about the Frauchiger\n Renner result that may have gone hithertofore unnoticed.
\end{abstract}

\maketitle

\section{Introduction}\label{secintro}
More than a century after its discovery, quantum mechanics arguably remains the most successful theory for the physical world at the fundamental scale. Yet, it is perhaps the physical theory that has most stubbornly eluded the grasp of human intuition. Amongst the enduring puzzles that quantum mechanics has posed at us, a certain category can be associated with the distinctive theme of \emph{quantum mechanics and intelligent agents}. The history of this category of quantum puzzles goes back to the pioneering Gedankenexperiment of ``Wigner's friend''~\cite{Wigner67}.

In more recent times, Frauchiger and Renner (hereafter ``\frp'') \cite{FR18} devised a Gedankenexperiment uncovering an inconsistency between quantum mechanics and certain seemingly innocuous commonsense reasoning principles. \frp's work has been seminal to a growing body of literature at least partly intended as a response thereto\m favourable at times and critical elsewhere, but altogether expanding our understanding. Much of this literature is devoted to finding alternative axiomatizations of quantum\hyp mechanical reasoning that do not suffer inconsistency in the face of \frp\hyp like arguments. Indeed, \frp\ themselves include in their work suggestions for ways to defuse the inconsistency within various interpretations\footnote{Unless specified otherwise, ``interpretation'' is here used in the semi\hyp technical sense of ``interpretation of quantum mechanics''.}.


Subsequent works have by and large accepted the soundness of \frp's main theorem \emph{as it stands}\m in particular, conditional on the manner in which the agents in the Gedankenexperiment reason. However, a common critique has been to question the rationality attributed to these agents, and to suggest that the inconsistency would be eliminated simply by having the agents reason ``truly rationally''. Of course, it has always been acknowledged, even by FR, that\m ipso facto their theorem\m\emph{the overall chain of reasoning by their fictional agent(s) is fallacious}. Rather, one of the central messages of their work (and of their subsequent responses to critiques) is that \emph{an exercise of localizing the fallacy to any proper part of the chain is (1) interpretation\hyp dependent in its details and, furthermore, (2) often not straightforward even within a single interpretation}.

In the present work, we propose a simple operational principle, which we call \emph{superpositional solipsism} (SPS), to inform inferential reasoning by quantum\hyp mechanical agents. Using the \frp\ Gedankenexperiment as a case study, we examine \ref{sps} in more detail and find that it provides a prescription for intersubjective inference that is (1) sound against \frp\hyp like arguments, (2) maximally predictive subject to this soundness, and (3) arguably interpretation\hyp agnostic in restricting to operational considerations. We highlight a few features that justify and recommend SPS\m some physical, some pragmatic, and some philosophical.

Before we proceed, we note some important points about the motivation, content, and scope of our work. Firstly, while it is inspired and motivated by the lively debate seeded by \frp, our intention is not to register any particular position in that debate. Rather, we seek\m and claim to have found in SPS\m operational principle(s) that may inform the reasoning of quantum\hyp mechanical agents in situations like those in the \frp\ Gedankenexperiment. Notably, SPS does not fit neatly into any of the distinctive roles that \frp's axioms (or substitutes thereof) play; rather, it functions as a sort of ``corrective'', to be applied in conjunction with other principles.

Secondly, we understand that SPS, and some of the arguments we make for it, may already have been put forth in some form in the literature. We believe, however, that the overall collection of ideas and arguments presented in this work is original and of value to the community, not least for our commitment to conceptual simplicity. We will contextually cite all precedent sources of similar ideas in our knowledge; we regret any inadvertent omissions.

Finally, for the purpose of this work, we commit to a sort of \emph{mind\n body nondualism}, whereby all thought and reasoning is assumed to be embodied in physical states. In this we follow \frp\ and related work. The status of our arguments in a dualist model may be an interesting problem for future work.

\section{The Frauchiger\n Renner theorem and subsequent literature}\label{secFR}
It will be helpful to set the context with an overview of \frp's work and the body of literature that it has inspired. After reviewing \frp's main theorem, we will sample a small selection of the subsequent literature, to get a glimpse of the variety of perspectives from which \frp's theorem has been analyzed. This section will then help motivate the need for a simplified operational perspective, which we hope to bring through our work.

\subsection{The FR framework}
At its core, \frp's work concerns three provisional axioms of quantum\hyp mechanical reasoning that could be employed by rational agents who are themselves governed by quantum mechanics\footnote{Here we will state the axioms in semi\hyp formal language, with no loss of essence from their original, formal formulations.}:
\begin{enumerate}
\item[{\crtcrossreflabel{(Q)}[axmq]}] If an agent has established that a quantum system S is in a pure state $\ket\psi$ at time $t$, then they may infer that a projective measurement of S at time $t$, with respect to an orthogonal basis $\left\{\ket{\psi_x}\right\}_{x=1}^d$ containing $\ket{\psi_1}=\ket\psi$, will yield the outcome $x=1$ with certainty. This is a weakened version of the Born rule for probability assignments to quantum measurement outcomes.
\item[{\crtcrossreflabel{(C)}[axmc]}] If an agent A has established that another rational agent B, with the same prior knowledge as A's, has established within the same physical theory as A's that $x=\xi$ at time $t$, then A may ``inherit'' the inference that $x=\xi$ at time $t$.
\item[{\crtcrossreflabel{(S)}[axms]}] If an agent has established that $x=\xi$ at time $t$, then they must necessarily deny that $x\ne\xi$ at time $t$.
\end{enumerate}

Based on an ingenious variation on the ``Wigner's friend'' Gedankenexperiment, \frp\ show the following.
\setcitestyle{authoryear}
\begin{theorem}[\citet{FR18}]\label{thm1}
Any theory that satisfies all of the axioms \ref{axmq}, \ref{axmc}, and \ref{axms} necessarily leads agents using it to contradictory inferences.
\end{theorem}
\setcitestyle{numbers}
We will discuss their detailed arguments in Section~\ref{seccas}, where we apply our own ideas to their Gedankenexperiment.

Having established their main theorem, \frp\ raised the question of which of the mutually\hyp inconsistent axioms were upheld by various common interpretations, summarizing their investigation of this question in Table 4 of their paper. In this connection, they also discussed what modifications could be made to those of the axioms that did not hold in a given interpretation. We colloquially refer to this avenue of research as \emph{the \frp\ framework}.

\subsection{The post\hyp FR intellectual landscape}
\frp's discussion with regard to the status of their provisional axioms in various extant interpretations has since been supplemented by others' work that has given detailed interpretation\hyp specific accounts \cite{Sud17,Drezet18,Sud19,LC19,DFS20,Kastner20,Bub20,Sud20}, addressed the issue in the language of formal logical systems~\cite{NdR18,Boge19}, examined the \frp\ theorem in connection with the philosophical underpinnings of science~\cite{Healey18}, and altogether advanced a vast variety of frontiers in light of the theorem.

This is but a very small sample of the literature in this area, and does not even consider the experimental work. Nevertheless, we find the discourse to be dominated by interpretation\hyp based accounts, and in places technically arcane. We therefore perceive a need for a treatment that combines bare\hyp bones operationality with a simplicity of formalism. We will expand on this in the next section.

\section{Towards a simple operational approach}\label{secop}
A burgeoning of diverse literature in a newly\hyp emergent avenue bodes well for the associated disciplines. Nevertheless, it could be unsettling to some\m it is to us\m that the body of literature inspired by \frp\ does not seem to be converging towards a scientific consensus on such questions as \frp's titular ``Can quantum theory consistently describe the use of itself?''. Besides, much of this literature studies axiomatic systems that may explicitly include interpretational details (such as ascribing definiteness to indirectly inferred events). Such study is undoubtedly of value in its own right. However, it may be argued that questions such as the one above are better addressed in an interpretation\hyp agnostic framework.

The bare formalism of quantum mechanics (henceforth ``quantum mechanics''), consisting of Schr\"odinger's equation together with Born's rule, is operationally testable and has stood the test. On the other hand, the various interpretations of quantum mechanics, which build upon the latter, have been operationally indistinguishable thus far. They may remain so at least for the foreseeable future for technological reasons, if not forever for fundamental reasons~\cite{CR11,BW18}.

In the present work we seek \emph{operational principle(s)} to help agents subject to quantum mechanics reason. We argue that such a prescription must be based only on the operationally\hyp relevant part\m namely, quantum mechanics\m of whatever quantum\hyp based theory the agents live and reason under. That is, in a world actually governed by any interpretation, an agent subscribing to any interpretation must be able to use the principle(s) soundly and intersubjectively with other agents subscribing to their own favourite interpretations. Of course, it is possible that quantum mechanics does not uniquely determine an operational reasoning framework; the principle we propose here is offered as merely one possible option.

Another issue we perceive in this connection is that many expositions employ technical machinery (modal logic, quantum histories, pilot waves, etc.)\ that are relatively unfamiliar to researchers in the mainstream of quantum information science, such as it is. The various formal representations of quantum mechanics, equivalent in their operational import, can all be described essentially interchangeably as ``the formalism of quantum mechanics''. However, the one we use here is arguably the de facto lingua franca in quantum information. In proposing an operational principle that looks simple in this language, we hope to contribute to the accessibility of the topic in this broad field.

In light of its distinctly interpretation\hyp agnostic motivation as laid out above, our work may be regarded as belonging not in the \frp\ framework proper, but in a closely\hyp allied avenue, albeit inspired by the former. It is almost certainly far from the first such work in the vast body of post\hyp \frp\ literature.

\section{The principle of superpositional solipsism}\label{secsol}
We are now ready to present our operational principle. We intend it as a corrective or amendment to be applied on every use of \frp's Axiom~\ref{axmq} or \ref{axmc}, in a manner that will become clear as we discuss. Focusing on operationally\hyp relevant elements of formalism, we do not concern ourselves with their Axiom~\ref{axms}.

Consider a primitive case where our prospective principle could be put to test: an agent Alice, inhabiting a quantum system A, is trying to reason in a situation involving a qubit system S. First, Alice uses a trusted device to prepare S in the pure state
\be
\ket\psi_\rS:=\fr1{\sqrt2}\left(\ket0_\rS+\ket1_\rS\right),
\ee
whose description is known to Alice. They then use another qubit, M, to perform a projective measurement on S in the $\left\{\ket0,\ket1\right\}$ basis, and then observe the outcome. To keep things simple, as in \cite{FR18}, we assume that Alice's state upon having observed either outcome is pure. Thus, the overall state of Alice and the two qubits at this point has the form\footnote{Although we use a many\hyp worlds\n like calculus where a measurement is modelled as an entangling unitary process involving the measurement device and the measured system, our description can be translated to a collapse\hyp based formalism, or indeed, any formal representation of quantum mechanics.}
\be\label{eqphis}
\ket\phi_{\rS\rM\rA}:=\fr1{\sqrt2}\left(\ket0_\rS\ket0_\rM\ket{\mathrm{zero}}_\rA+\ket1_\rS\ket1_\rM\ket{\mathrm{one}}_\rA\right).
\ee
As with Wigner's friend, suppose it were agreed beforehand between Alice and their friend Bob that the latter was to now perform a (previously agreed\hyp upon) projective measurement on the composite system SMA.

How must Alice reason about the outcome of the impending measurement? An obvious option would be to simply apply Born's rule, e.g.\ as formulated in \frp's Axiom~\ref{axmq}\footnote{The validity of \ref{axmq} is conditional on implicit physical and metaphysical assumptions, an assortment of which may be found in \cite{FR18,NdR18,Sud19}. Of course, no such assortment can be deemed definitively complete; in this work we do not contest these implicit assumptions.}, on the overall experimental design. This would, of course, yield operationally\hyp sound and maximally\hyp informative predictions from an apriori perspective on the experiment, namely that the outcome is completely random, with each of the two possible values occurring with a probability 1/2.

However, one would hope that Alice's special epistemic circumstance in having observed a specific outcome should give them some predictive edge over the apriori perspective. To inform such \emph{situational reasoning}, we introduce our principle of \emph{superpositional solipsism} (SPS):
\begin{axm}[{\crtcrossreflabel{\textnormal{SPS}}[sps]}]
An agent must always reason based on the assumption that their current observational state is definite and not superposed or mixed with other potential states, which the agent does not currently experience.
\end{axm}
\begin{cls}[Observational state]
The ``observational state'' of an agent includes all definite values that the agent has determined for observables external to the agent through all the measurements they have performed thus far, as well as all mental / internal representations of such definite values, that the agent knows they possess.
\end{cls}
\begin{note}
While we extend the scope of \ref{sps} to disregard both coherent superpositions and incoherent mixtures in agents' reasoning, the former has a more important influence on the outcomes of rational inferences; hence the name we have chosen for the principle.
\end{note}

In the remainder, we will continue to state such clauses to clarify \ref{sps}. Coming back to the example at hand, Alice's observational state notably includes the piece of classical information ``zero'' or ``one'' concerning the measurement outcome they observed. Here, one might wonder, what if we express the very same state $\ket\phi_{\rS\rM\rA}$ as
\begin{widetext}
\be
\ket\phi_{\rS\rM\rA}=\fr12\left[\ket{\Phi^+}_{\rS\rM}\left(\ket{\mathrm{zero}}_\rA+\ket{\mathrm{one}}_\rA\right)+\ket{\Phi^-}_{\rS\rM}\left(\ket{\mathrm{zero}}_\rA-\ket{\mathrm{one}}_\rA\right)\right],
\ee
\end{widetext}
where $\ket{\Phi^\pm}_{\rS\rM}:=\fr1{\sqrt2}\left(\ket0_\rS\ket0_\rM\pm\ket1_\rS\ket1_\rM\right)$? Could one not argue then that the states $\fr1{\sqrt2}\left(\ket{\mathrm{zero}}_\rA\pm\ket{\mathrm{one}}_\rA\right)$ are equally valid as ``definite observational states'', indicating the ``observation'' of $\ket{\Phi^\pm}_{\rS\rM}$ on SM?

This is where the clause ``\emph{that the agent knows they possess}'' is important: for agents like us humans, observational states include elements of self\hyp awareness that comprise redundant copies of certain pieces of classical information such as the value ``zero''. More generally, in all agents that have an internal mechanism to multiply (necessarily classical) information into redundant copies, one can define the definite observational states as those that correspond to definite values of such copiable information. An instance of the application of \ref{sps} can then be described as a formal decoherence with respect to the ``preferred'' or ``classical'' basis of the agent's internal states (by ``formal'' we mean that this decoherence only needs to be applied by the agent in their mental representation of the situation).

For more general agents (e.g.\ quantum computers), which fundamentally lack fixed preferred bases, it may still be possible to define an effective classical basis in a particular situation where the agent is to make an inference. We will expand on this (and on general POVM\hyp based and weak measurements) later in Section~\ref{qag}; for now, we will take as given that any specific instance where \ref{sps} is to be applied, i.e.\ where an agent is to make inferences, comes with a well\hyp defined notion of definite observational states. Importantly, the framework in which \frp's Gedankenexperiment is cast implicitly contains such a notion, and even so, the sort of questions considered by \frp\ and subsequent literature are evidently nontrivial to answer.

Where does all this lead our Alice? Let us suppose that Bob's measurement is relative to a basis that contains the vectors $\ket0_\rS\ket0_\rM\ket{\mathrm{zero}}_\rA$ and $\ket1_\rS\ket1_\rM\ket{\mathrm{one}}_\rA$. Based on \ref{sps}, Alice would predict that Bob would certainly get the same outcome as Alice. This prediction would, of course, never go wrong in this simple case. Thus, in this case, Alice would not reason unsoundly in ignoring the superposition; on the contrary, they earn an operational advantage by applying \ref{sps} instead of taking an apriori perspective on the experiment.

But what about cases where Bob is to make a different measurement? Surely, in these cases Alice would be misled by disregarding the superposition?
\begin{obsm}[Operational immunity of predictive power]\label{obpre}
If Alice is a wilful subject in an experiment that by design depends on a superposition of certain predetermined definite observational states of theirs, or that otherwise involves absolute control of their cognitive apparatus by Bob, then
\begin{enumerate}
    \item Alice effectively cedes all rational agency for the duration of the relevant coherent part of the experiment.
    \item\label{notinf} Even if the agents do prescribe an inferential system for Alice, any instance in the experiment of such a system's situational application is necessarily scripted into the experiment's design, and as such, plays no role as an inferential aid.
    \item\label{script} In order to script such inference into the experimental design, the agents would anyway have had to execute the inference's line of reasoning in advance, obtaining then and there whatever insight was to be gained from it; in this prior reasoning, they could employ quantum mechanics in its full predictive power.
\end{enumerate}
In particular, enforcing \ref{sps} on Alice's situational reasoning would result in no loss of predictive power.
\end{obsm}
It is worth while to expand on points \ref{notinf} and \ref{script} above to preempt some potential misunderstandings. Firstly, in response to our claim that a pre\hyp scripted inference does not function as an inferential aid, one might object thus: In the deterministic (i.e.\ not stochastic) classical computing model, the hardware and program together completely determine the outcome of an instance of computation. In this sense, the outcome can be considered to have been ``present all along'' in the hardware\n program composite, existing as a foregone conclusion regardless of whether an actual process of computation is carried out. Would we then also deem such a process as functionally not an inferential aid, thereby summarily trivializing the value of deterministic classical computation?

Not so: we would in fact draw an important distinction between deterministic classical computation and the kind of ``scripting'' mentioned in point~\ref{notinf}. In the former model, one may write a program to, say, compute the first ten digits of the decimal expansion of $\pi$. In doing so, one is rationally certain that executing the program will result in a predetermined outcome, and that this outcome will coincide with the formal definition of ``the desired digits of $\pi$'' that one can accept without actually knowing the digits. Crucially, one does not need to \emph{have explicitly computed the outcome in advance} by other means in order to write the program in the first place; if they did, then the program would indeed be functionally useless.

On the other hand, in an experiment like Alice and Bob's above, or like the \frp\ Gedankenexperiment, which we will discuss in Section~\ref{seccas}, measurements with respect to superpositions of definite observational states are part of the design. Engineering such measurements, in turn, requires the agents to know the relevant observational states in detail. Quantum mechanics forbids one to determine the detailed state of a system through any single observation\footnote{Moreover, even a single observation, besides divulging only a small amount of information about the state of the system, pushes the system out of the state it was in before the measurement.}, and so e.g.\ Bob couldn't just wait for Alice to get into the entangled state $\ket\phi_{\rS\rM\rA}$ of Eq.~\ref{eqphis} and then, upon their first encounter with this state, manage to perform a measurement with respect to it or another superposed state. They would have to first get Alice into the exact same state a large number of times, perform various measurements, and thereby learn the details of Alice's definite observational states to a sufficient degree of precision. Only then would they be able to perform the measurement desired in the actual planned experiment.

Clearly, in this process Bob would already have had occasion to note the outcomes of Alice's inferences in the prospective actual experiment. This is what we alluded to in point~\ref{script}, and thus the sort of inferences made by agents in Alice's situation are \emph{truly operationally useless}: not because the outcomes are predetermined by the setup, but because \emph{one needs to know the outcomes in advance to even build the setup}.

Another interesting point in this connection, which has received some attention in \cite{FR18,NdR18}, is the following: if Alice was rational in their considerations before agreeing to subject themselves to such an experiment, they would also be wise to the operational unreliability of any memory they will later retain of the experiment\m even ones of inferences they made during the experiment. Therefore, should Alice remember making in\hyp experiment situational inferences that since turned out to be refuted by the eventual experimental outcome, they have no reason to be embarrassed or charged with fallacious reasoning (unless they insist on taking their memories seriously); we will revisit this point in Section~\ref{secop2}.

Yet, we emphasize that the unreliability of memories of such an inferential action is not the reason the action itself must not be considered a reliable inference: in view of the ``scripting'' arguments above, it is not the inference's \emph{reliability} that is in question, but rather its very ``\emph{inferenceness}''. For contrast, consider an agent who is in a position to make a sound situational inference and also apply it situationally, in true possession of their agency. Say the agent also knows that they will later lose the fidelity of their memory of the inference. In this case, the inference could be deemed unreliable outside of the situation, owing to the unreliability of its memory.

However, in cases where cognitive volition itself is absent, and where moreover the outcomes of ostensible inferences are known in advance, such actions are not to be deemed reliable inferences, simply because \emph{they aren't even inferences}. In fact, experiments that rely on such actions rob the subordinate agents not only of the freedom to reason, but also of the \emph{freedom to abstain from reasoning}! As such, these actions are not worthy of scrutiny with regard to their soundness as inferences. Hence, echoing the concluding remark of Observation~\ref{obpre}, applying \ref{sps} does not compromise the predictive power of actions that can truly operationally be considered inferences.

This is almost surprisingly reassuring in the face of the apparent radical reductionism of \ref{sps}. But what recourse do agents have, if they wish to retain \emph{true} situational rational agency during such an experiment?
\begin{obsm}[Pragmatism regarding situational rational agency]\label{obag}
If Alice retains any part of their cognitive apparatus, however small, outside of Bob's control and employs it in volitional situational reasoning, they invalidate a commensurate coherent element of the experiment by that very act, and consequently don't have to take this element into consideration for their reasoning.
\end{obsm}
\frp\ do note that their paradoxical prediction would fail if any information leaked from the coherent control of their Gedankenexperiment's superagents; this point has since been examined in more detail in later work~\cite{ELM+20}. However, the need to engineer (as part of the experimental design) specific inferential chains into the cognitive machinery of the measured agents has either been implicitly assumed, or else escaped the authors' notice; we will expand on this in Section~\ref{seccas}.

Here, it is important to note that we need make no assumptions of the agents' being macroscopic or complex, their observational states' being technologically infeasible to superpose, etc. Observation~\ref{obag} becomes relevant even in cases where an agent retains agency over \emph{so much as but a single qubit} of their cognitive faculty outside of the controlled experiment, however large and complex the controlled part may be.

Failing such retention of agency, an ostensibly inference\hyp like action that is, for whatever reason, scripted into the experimental design \emph{might as well obey} \ref{sps}. In fact, we take a more radical stance and argue that \ref{sps} would only be the pragmatic way even to script inferences: in the event that an agent strays from the script\m inadvertently, mutinously, or for other unforeseen reasons\m a commitment to \ref{sps} at least ensures that the decoherence resulting from the transgression coincides with the \ref{sps}\hyp based formal decoherence in the agent's reasoning, and therefore that the information that gets unplannedly leaked acts in favour of validating any situational inference the agent makes. We will make this argument more concrete in Section~\ref{seccas}, but in anticipation of it, we have the following.
\begin{cls}[Overriding superpositional design]\label{overr}
Where there is potential conflict, an agent must in their reasoning uphold \ref{sps} over superpositional elements that they know to be part of the experimental design.
\end{cls}
Sometimes, an agent could be subject to an experiment that involves measurements with respect to \emph{some} superposed observational states but does not prescribe (or engineer) any particular observational states on the agent; indeed, such experiments are already being conducted, with the aim of advancing our understanding of human cognition. In such cases, \ref{sps} is even more self\hyp evidently applicable than in the aforementioned scenarios where specific observational states are required by design.

What about cases where a ``malevolent superagent'' operates on an agent's cognitive machinery without the latter's knowledge?
\begin{obsm}[Pragmatism regarding unfalsifiable branches]
The hypothesis that the wavefunction of an agent's world contains, in addition to a branch corresponding to the agent's current definite observational state, a superposition with other branches, in a manner unknown to or unanticipated by the agent, is fundamentally unfalsifiable. Therefore, the agent would only be pragmatic in ascribing no credence to this hypothesis.
\end{obsm}
The operational standing of such a hypothesis is spiritually akin to that of the hypothesis that one is a Boltzmann brain, or that one lives in a superdeterministic world. Pragmatism against such hypotheses has been supported on rigorous grounds in recent work by M\"uller \cite{Mueller20}, as part of a radically solipsistic approach to physics. Indeed, we humans have been practicing pragmatism against the hypothesis of unknown branches of the wavefunction throughout the history of our use of quantum mechanics; we could scarcely do without it. What is notable here is that \ref{sps} effectively extends this attitude even to \emph{branches that we expect to be present but do not currently experience}.

To summarize this section: we introduced the principle of superpositional solipsism (\ref{sps}), which broadly prescribes that rational agents commit to a ``formal collapse'' to their current observational state for the purpose of drawing inferences. We showed that \ref{sps} can be applied in conjunction with \frp's Axiom~\ref{axmq} without compromising agents' predictive power, but on the contrary sometimes \emph{improving} it. We also elucidated the operational simplicity of \ref{sps} in not prescribing different courses of action in the face of diverse forms of observational superposition\m with known or unknown superagents, and in the presence or absence of specification in what the measured agents are required to do.

Moving on, we must investigate how \ref{sps} works together with Axiom~\ref{axmc}, which concerns situations where agents reason from their mental reconstructions of other agents' perspectives. For a complete treatment of this aspect, we will need more complex situations than what the Alice\n Bob experiment above affords. In the upcoming section we will use \frp's Gedankenexperiment as our test case to examine \ref{axmc} as well as to uncover a few more interesting implications of \ref{sps}.

\section{A deeper investigation of the principle}\label{seccas}
Building on the previous section's insights into the principle of superpositional solipsism (\ref{sps}), here we will probe the principle more thoroughly. To this end, it will be helpful to apply it to \frp's Gedankenexperiment, which features situations more complex than the simple Alice\n Bob Gedankenexperiment above.

\frp's Gedankenexperiment, an extension of ``Wigner's friend'', involves four agents, whom they call F, $\fb$, W, and $\wb$\m two friends and, presumably, two Wigners (Margit and Eugene, say). We provide a detailed description of the Gedankenexperiment in Appendix~\ref{secFR}. Here, we proceed by first summarizing how \frp\ use it to derive their main result, Theorem~\ref{thm1}. Thereafter, we will scrutinize their arguments through the lens of our \ref{sps} principle. This will enable us to (1) identify certain critical fallacies in the reasoning of the hypothetical agents; (2) illustrate how to apply \ref{sps} in conjunction with \frp's Axiom~\ref{axmc}; and (3) afford some more insight into the operational implications of \ref{sps}.

\subsection{FR's arguments}\label{frarg}
Upon some examination of the experiment, one finds that $\wb$ has a nonzero probability of obtaining the outcome $\overline w=\overline{\mathrm{ok}}$ in any given round. Consider an instance where they do obtain this outcome, and say they then wish to predict the outcome $w$ of W's subsequent measurement. The form of the overall state of all relevant systems at this point in the experiment, Eq.~\eqref{eqxi}, shows that the outcome $\overline w=\overline{\mathrm{ok}}$ is exclusive of the outcome $z=-\fr12$ of F's measurement. Therefore, $\wb$ may reason that F must have obtained $z=\fr12$.

At this juncture, one straightforward line of reasoning $\wb$ can take is to invoke Axiom~\ref{axmq} and infer that, in the next step of the experiment, W will obtain outcome $w={}$ok with probability $1/2$; this inference is indeed vindicated from our detached point of view, which can be seen as an application of \ref{axmq} to the overall experiment. However, the stated axioms leave other lines of reasoning open to $\wb$. For example, they may put themself in F's shoes and imagine what the latter would have rationally inferred from $z=\fr12$ (which by now is established from $\wb$'s perspective). They may then invoke \ref{axmc} to inherit F's presumed inference. The question, then, is: how would F have reasoned upon getting the outcome $z=\fr12$?

F, too, could just invoke \ref{axmq} on their own lab to infer (what we know to be) the correct expected outcome of the experiment. Alternately, from the form \eqref{eqphi} of the state after their measurement, F could deduce that $z=\fr12$ is inconsistent with $r={}$heads, and that, therefore, $r$ must be tails. They may then proceed to invoke \ref{axmc} to reason on $\fb$'s behalf in light of $r={}$tails.

What would $\fb$ have concluded from this outcome? Based on the form \eqref{eqpsi} of the state post $\fb$'s measurement, a straightforward application of \ref{axmq} would lead them to predict that $w={}$fail with certainty. Extracting this hypothetical inference through the two nested layers of \ref{axmc}, $\wb$ may thus predict that $w={}$fail with certainty. This prediction, however, is in conflict with $\wb$'s conclusion from reasoning by \ref{axmq} alone, which admits the nonzero probability $1/2$ for $w={}$ok to occur; at any rate, the certain\hyp fail prediction is bound to eventually be manifestly invalidated on some round of the experiment.

This already appears to establish inconsistency in any axiomatic system that includes \ref{axmq} and \ref{axmc} (note that Table 4 of \frp\ does not provide evidence that any theory can really uphold even these two axioms simultaneously). FR, however, reason that $\wb$ and their friends may restore consistency by admitting the possibility for a quantity like $w$ to take multiple values simultaneously.

As an aside, we remark that even in theories that violate \ref{axms}, consistency with \ref{axmq} would, to the best of our knowledge, preclude the ``many simultaneous values'' construct from helping in cases where quantum mechanics predicts one value \emph{with certainty}. Nevertheless, to close this presumed loophole, \frp\ also explicate Axiom~\ref{axms} and direct their main theorem only at axiomatic systems that include all three axioms.

Having thus seen how \frp\ argue for Theorem~\ref{thm1}, we will now apply our principle to their Gedankenexperiment.

\subsection{An operational assessment of FR}\label{opfr}
We first note that \frp's outermost narrative does not disclose whether, or in what manner, $\fb$ and F reason. It only discusses how $\wb$ reasons; as part of this reasoning, $\wb$ imagines how F might reason. Even further out, \frp's description of $\fb$'s reasoning is part of the imagined narrative of $\wb$'s imagined F. Setting aside this nested structure for now, we will first scrutinize each agent's reasoning chain as though it took place in \frp's outermost narrative.

Let us start with $\fb$, who takes the simplest line of reasoning: $\fb$ reasons, upon getting the outcome ``tails'' and preparing S in the state $\ket\rightarrow_\rS$, that the state of L just before W's eventual measurement will be $\ket{\mathrm{fail}}_\rL$. Based on this, $\fb$ concludes that W will obtain the outcome $w={}$ok. In this case, $\fb$ actually adheres to \ref{sps} in disregarding the branch where they obtained the outcome ``heads'' in the state $\ket{\psi}_{\rR\db\fb\rS}$ [Eq.~\eqref{eqpsi}].

In the language of Section~\ref{secsol}, however, $\fb$'s reasoning action (if any) is scripted into the experimental design\m they are, after all, to be measured later by $\wb$. As such, any judgement of the soundness of this action as an inference is operationally meaningless; for now we will content ourselves with the observation that the inference adheres to \ref{sps}, but later we will explore how stronger judgements could be made.

Moving on to the slightly more complex inferential chain imputed to F, the latter's first relevant inference is upon seeing the outcome $z=\fr12$ to their measurement: they reason that this outcome, being incompatible with $r={}$heads, signals that $\fb$ must have obtained $r={}$tails. In this step F, too, upholds \ref{sps}. They then invoke Axiom~\ref{axmc} to ``summon their inner $\fb$'', whom they imagine in the state of having obtained $r={}$tails.

$\fb$ reasons, as noted above, on the basis of the outermost experimental design. In particular, as one of the steps in their chain of reasoning, $\fb$ infers that the state of L just before W's measurement will be $\ket{\mathrm{fail}}_\rL$. This inference would adhere to \ref{sps} if the \emph{real} $\fb$ situationally carried it out. However, such an inference on the part of the \emph{figmental} $\fb$ in real F's imagination would run afoul of F's commitment to \ref{sps}: it would result in F considering themselves to be in the state $\ket{\mathrm{fail}}_\rL$, a superposition that includes observational states other than the one they currently experience. After this step, $\fb$ bases all of their subsequent reasoning on the assumption that they are in the superposition state. Again, the status of F's action as a bonafide inference\m sound or fallacious\m aside, we note that their overall line of reasoning violates \ref{sps}.

Moving on to $\wb$: their probabilistic prediction for $w$ based solely on \ref{axmq} applied to their current observational state upon seeing $\overline w=\overline{\mathrm{ok}}$, as well as their inference that $z=\fr12$, adheres to \ref{sps}. However, their reasoning chain based on $z=\fr12$ and the nested applications of \ref{axmc} violates \ref{sps}. To wit, when $\wb$ inherits their imagined F's inference that $r={}$tails, they must carefully consider the implications: reasoning based on a commitment to $r={}$tails would cause them to ascribe to themselves the state $\fr1{\sqrt2}\left(\ket{\overline{\mathbf{fail}}}_{\lb\eb\wb}-\ket{\overline{\mathbf{ok}}}_{\lb\eb\wb}\right)$, thereby positing branches whose observational states they don't currently experience.

Unlike $\fb$ and F, $\wb$ does have the freedom to reason as they please, without affecting the outcome of the experiment. $\wb$'s reasoning can therefore be judged on its merit, and we see that it is both unsound (as found by \frp) and in contradiction with \ref{sps} (as we found).

Of the diverse possible ways to ``localize the fallacy'' in the agents' reasoning, an \ref{sps}\hyp based analysis localizes it to $\wb$'s inference that $r={}$tails. Thus, while their previous inference $z=\fr12$ is vindicated, as is their \ref{axmq}\hyp based probabilistic prediction for $w$, their line of reasoning from $r={}$tails onwards rests on flawed grounds. In addition, if $\wb$'s imagined inferences on behalf of F and $\fb$ are actually scripted as such into the experimental design, \ref{sps} would also find fault in F's reasoning. We merely note this in passing, and do not believe that the ``flaw\hyp localizing'' exercise has any operational significance in general.

We saw some indications above that intersubjective reasoning informed by \ref{axmc} can conflict with \ref{sps}. In order to retain as much as possible of the power of \ref{axmc}, we add the following clause.
\begin{cls}[Intersubjective inference]\label{intinf}
When using Axiom~\ref{axmc} for intersubjective inference, an agent must add the definiteness of their own current observational state to the premises underlying the imagined reasoning of all other agents.
\end{cls}
For example, if the real F invokes \ref{axmc} to reason on $\fb$'s behalf, they must imagine that $\fb$ ascribes to L the state $\ket{\tfrac12}_\rL$. This way, they would not inherit the real $\fb$'s prediction ``$w={}$fail''. Upon close examination, one finds that \ref{axmc} is really no more helpful than \ref{axmq}: the latter could just as well be used by an agent to infer (based on their knowledge of the experimental design) the values of unobserved observables from definite values of observables they have already determined (i.e.\ contents of their current definite observational state). The definite values constituting their premise may be determiners of the states of external systems; for the purpose of such inference, there is no reason to distinguish between external systems with and without agency.

But if an agent should consider how another agent might have reasoned, Clause~\ref{intinf} of \ref{sps} effectively exhorts them to uphold their own perspective over the other's in the event of a conflict; in this sense, Clause~\ref{intinf} is just a special case of Clause~\ref{overr}. Based on this prescription, an agent might think along the lines of ``My friend would certainly have inferred $x=\xi$ based on their perspective, and they were sound in reasoning so; however, if my perspective informs me that $x\ne\xi$, I must uphold the latter.'' In a generic instance, inferences drawn by applying \ref{axmc} on another agent's independent perspective would include ones that do not conflict with one's own definite observational state (e.g.\ $\wb$'s use of \ref{axmq} to predict the probabilistic $w$ outcome from their previous inference that F must have seen $z=\fr12$). Such inferences would, of course, remain valid under the constraints imposed by \ref{sps}.

\subsection{Virtual disclosure and transmissive stability}\label{secvd}
In classical mechanics, virtual displacement and virtual work are formal abstractions that can be used to derive equations that govern the actual dynamics of physical systems. Here, we will propose a similar abstraction\m\emph{virtual disclosure}\m to help in the operational assessment of inferences.
\begin{clm}[Virtual disclosure]\label{vdis}
Consider an experiment featuring an agent who, with prior knowledge of the detailed experimental design, performs an inference\hyp like action. Suppose that, just after the agent performed this action, the experimental design were temporarily violated by copying the agent's observational state onto a system external to the experiment. Suppose further that the experiment were otherwise carried out as designed. Then,
\begin{enumerate}
    \item If the action was not already a true operational inference in the sense of Section~\ref{secsol}, it is elevated to that status by the (unplanned) copying action.
    \item The resulting inference is sound if and only if it adheres to \ref{sps}.
\end{enumerate}
\end{clm}
We could summarize this \emph{principle of virtual disclosure} as follows: ``An inference\hyp like action that is not a true operational inference adheres to \ref{sps} if and only if it is elevated to a sound operational inference by a virtual disclosure.'' Note that the ``copying'' embodying the virtual disclosure is defined relative to the relevant ``classical basis'' or ``basis of definite observational states'', and therefore results in a decoherence relative to this basis when only systems internal to the experiment are considered.

We will leave the formalization and rigorous proof of this claim (including the formalization of such terms as ``true operational inference'') for future work, but here present intuitive arguments for it. The case where the inference concerns something unrelated to the experiment is uninteresting and trivial. Moving on to the case where the inference pertains to the experiment and its outcomes, there are two salient sub\hyp cases: (1) the inferential action has no influence on the experiment's outcome; (2) the action does influence the experiment's outcome. In case (1), the experiment cannot possibly be one involving the superposition of different observational states of the agent; therefore, the agent would reason soundly by committing to the definite values they have determined for observables.

In case (2), if the action's influence on the experiment's outcome was through a coherent superposition with other observational states, then (as noted above) the disclosure would break the requisite coherence. In the resulting decohered version of the experiment, if no other design feature were violated, then again the agent would reason soundly based on the definite values they observed. This shows the ``if'' part. We will provide evidence for the ``only if'' direction presently.

If the action impinged on the outcome in a different way, not involving coherent superposition, then \ref{sps} per se\m whose dictate solely concerns unobserved superpositions\m is not likely to preclude a proper consideration of the relevant premises. We therefore expect the claim to be valid even for this case, but confirming this would require careful formalization. One category of problematic cases is of inferences whose conclusions are self\hyp referential, e.g.\ ``This statement is a secret between Alice and Bob'', ``This statement has been uttered an even number of times'', etc.

We could try to exclude such problematic cases from the scope of our claim by restricting it to inferences whose conclusions have the property that their truth value is invariant when they are transmitted or copied from one system to another. This property, which may be called \emph{transmissive stability}, could possibly be used (after appropriate formalization) as a definition for ``true operational inference''. The science of formal languages and logical systems might already have formalizations of such notions, and we hope to explore relevant connections in future work.

As promised above, we will now present some arguments for the ``only if'' part of Claim~\ref{vdis} in the case where the inferential action influences the experimental outcome for reasons related to coherent superposition. The \frp\ Gedankenexperiment contains just such instances, and so we will base our arguments on these.

We have already seen that $\fb$'s reasoning adheres to \ref{sps}. As such, it would not help us probe the ``only if'' direction. Nevertheless, it is worth while to apply the virtual disclosure principle to it as a test case of the ``if'' part. Suppose that $\fb$ obtained $r$={}tails, prepared S in the state $\ket\rightarrow_\rS$, and then made the following inferences:
\begin{enumerate}
    \item S is in the state $\ket\rightarrow_\rS$; therefore the state of L just before W's measurement will be $\ket{\mathrm{fail}}_\rL$.
    \item Owing to the above state of L, W will certainly obtain $w={}$fail upon their measurement.
\end{enumerate}
A disclosure after either of these steps can effectively be modelled by preceding $\wb$'s measurement by the following isometry:
\begin{align}
    \ket{\hb}_{\lb}&\mapsto\ket0_\rG\ket{\hb}_{\lb};\nonumber\\
    \ket{\tb}_{\lb}&\mapsto\ket1_\rG\ket{\tb}_{\lb},
\end{align}
where G is a qubit external to the experiment. Following the rest of the experiment as per design, we eventually reach the state
\begin{widetext}
\be
\sqrt{\fr1{12}}\ket0_\rG\left(\ket{\overline{\mathbf{ok}}}_{\lb\eb\wb}+\ket{\overline{\mathbf{fail}}}_{\lb\eb\wb}\right)\left(\ket{\mathbf{ok}}_{\rL\rE\rW}+\ket{\mathbf{fail}}_{\rL\rE\rW}\right)+\sqrt{\fr13}\ket1_\rG\left(\ket{\overline{\mathbf{fail}}}_{\lb\eb\wb}-\ket{\overline{\mathbf{ok}}}_{\lb\eb\wb}\right)\ket{\mathbf{fail}}_{\rL\rE\rW}.
\ee
\end{widetext}
The external flag $\ket1_\rG$, which signals $\fb$'s inferences leading to the conclusion ``$w={}$fail with certainty'', does in fact necessitate the actual outcome $w={}$fail. Thus, \ref{sps}\hyp adherence renders $\fb$'s reasoning sound under virtual disclosure. Note also that $\fb$'s disclosed conclusion possesses transmissive stability and could, therefore, be broadcast indefinitely without affecting its factuality.

Moving on to the interesting case of F's reasoning, the latter consists of the following inferences (upon getting the outcome $z=\fr12$):
\begin{enumerate}
    \item $z=\fr12$; therefore $\fb$ must certainly have observed $r={}$tails.
    \item $\fb$ has observed $r={}$tails; therefore, they must certainly have concluded, ``W will certainly obtain $w={}$fail upon their measurement.''
    \item $\fb$ reasoned soundly in predicting so; by \ref{axmc}, I must predict likewise; therefore, W will certainly obtain $w={}$fail upon their measurement.
\end{enumerate}
As we did with $\fb$, let us virtually disclose F's inference by preceding W's measurement by the isometry
\begin{align}
    \ket{-\tfrac12}_{\rL}&\mapsto\ket0_\rG\ket{-\tfrac12}_{\rL};\nonumber\\
    \ket{\tfrac12}_{\rL}&\mapsto\ket1_\rG\ket{\tfrac12}_{\rL}.
\end{align}
Playing out the rest of the original experiment, we have the eventual state
\begin{widetext}
\be
\sqrt{\fr13}\ket0_\rG\ket{\overline{\mathbf{fail}}}_{\lb\eb\wb}\left(\ket{\mathbf{ok}}_{\rL\rE\rW}+\ket{\mathbf{fail}}_{\rL\rE\rW}\right)+\sqrt{\fr1{12}}\ket1_\rG\left(\ket{\overline{\mathbf{ok}}}_{\lb\eb\wb}-\ket{\overline{\mathbf{fail}}}_{\lb\eb\wb}\right)\left(\ket{\mathbf{ok}}_{\rL\rE\rW}-\ket{\mathbf{fail}}_{\rL\rE\rW}\right).
\ee
\end{widetext}
In this case, the virtual disclosure flag $\ket1_\rG$ signaling F's inferences fails to uphold their final conclusion. It is instructive to consider what would happen if $\wb$ had measured in the $\left\{\ket{\hb}_{\lb},\ket{\tb}_{\lb}\right\}$ basis instead: the virtual disclosure would in fact then vindicate F's first two inferences, although the final one would still fail. Of course, this happens to be the one inference that conflicts with \ref{sps}, which the first two adhere to.

Thus, the virtual disclosure exercise applied on F's reasoning chain provides support for both the ``if'' and the ``only if'' directions of Claim~\ref{vdis}. Again, all of F's inferred statements (including the false one) possess transmissive stability: their truth values remain invariant under transmission.

As for $\wb$, their reasoning does not need to be examined by virtual disclosure\m it is subject to \emph{actual} disclosure in the outermost, operational layer of the experiment, and all its inferred statements are transmissively stable. We have already seen that exactly those of their inferences that adhere to \ref{sps} are borne out by empirical evidence.

In Section~\ref{secsol}, we used the simple Alice\n Bob Gedankenexperiment to illustrate that \ref{sps}\hyp based reasoning can make more informative predictions than reasoning based on superposed observational states that are not directly experienced. In this section, based on the more complex \frp\ Gedankenexperiment, we have presented arguments supporting the stronger Claim~\ref{vdis}: if true, this would mean that neglecting \ref{sps} could lead not only to possible diminution in predictive power, but also to the loss of soundness.

\subsection{No-signalling and operational completeness of the wavefunction}
The \frp\ Gedankenexperiment exhibits some more curious quirks of quantum\hyp mechanical reasoning that are easy to miss at first glance. One of these concerns operations that commute under the formalism of quantum mechanics.

Recall our critical assessment of $\wb$'s reasoning in Section~\ref{opfr}, where we argued that their assignment of the state $\ket{\tb}_{\lb}$ to $\fb$'s lab would lead them to the \ref{sps}\hyp violating assignment $\fr1{\sqrt2}\left(\ket{\overline{\mathbf{fail}}}_{\lb\eb\wb}-\ket{\overline{\mathbf{ok}}}_{\lb\eb\wb}\right)$ to their own state. The reasoning chain makes its way ostensibly via an inference regarding F's lab L. Notice, however, that $\wb$'s measurement of $\lb$ commutes with F's actions. As far as quantum mechanics is concerned, $\wb$ might as well be making their inferences independently of what transpires on the L side of the story. With this perspective, we see plainly that $\wb$'s inference amounts to the absurdity ``My measurement found $\lb$ to be in the state $\ket{\overline{\mathrm{ok}}}_{\lb}$; therefore, $\lb$ must have been in the state $\ket{\tb}_{\lb}$.''

This idea of commuting operations can be extended to uncover closely\hyp related fallacies~\cite{SDDS20} in \ref{axmc}\hyp based reasoning:
\begin{obsm}[Operational fallacies in reasoning based on \ref{axmc}]
The state $\ket{\phi}_{\lb\rL}$ [Eq.~\eqref{eqphi}] of $\lb\equiv\rR\db\fb$ and $\rL\equiv\rS\rD\rF$ can also be prepared through the following steps:
\begin{enumerate}
    \item $\rF$ first prepares $\rS$ in the state $\sqrt{\fr13}\ket\uparrow_\rS+\sqrt{\fr23}\ket\downarrow_\rS$.
    \item $\rF$ then uses $\rD$ to measure $\rS$ in the $\left\{\ket\uparrow_\rS,\ket\downarrow_\rS\right\}$ basis, and makes a mental note of the outcome.
    \item If the outcome is $\uparrow$, $\rF$ prepares $\rR$ in the state $\ket{\mathrm{tails}}_\rR$; if the outcome is $\downarrow$, $\rF$ prepares $\rR$ in the state $\sqrt{\fr12}\left(\ket{\mathrm{heads}}_\rR+\ket{\mathrm{tails}}_\rR\right)$.
    \item $\rF$ sends $\rR$ to $\fb$.
    \item $\fb$ uses $\db$ to measure $\rR$ in the $\left\{\ket{\mathrm{heads}}_\rR,\ket{\mathrm{tails}}_\rR\right\}$ basis, and makes a mental note of the outcome.
\end{enumerate}
Overall, this preparation can be summarized through the rendering $\ket{\phi}_{\lb\rL}=\sqrt{\fr13}\ket{\tb}_{\lb}\ket{\fr12}_\rL+\sqrt{\fr23}\ket{\overline{\mathrm{fail}}}_{\lb}\ket{-\fr12}_\rL$ of the same state. In this case, $\fb$ would infer from $\ket{\fr12}_\rL$, predictively, that $\lb$ will later be in the state $\ket{\tb}_{\lb}$, as they do \emph{retrodictively} in the original Gedankenexperiment. But what would they then proceed to predict, based on \ref{axmc} applied on the future $\fb$? They are now faced with a quandary:
\begin{enumerate}
    \item If $\rF$ predicts that their own lab will later be in the state $\ket{\mathrm{fail}}_\rL$, then they must admit that $\fb$'s action has an influence on $\rL$ that violates the no\hyp signalling principle.
    \item Alternately, if $\rF$ predicts that their own lab will remain in the state $\ket{\fr12}_\rL$, then they must admit that \ref{axmc} prescribes different lines of reasoning from the same state, based on different preparations.
\end{enumerate}
\end{obsm}
On the other hand, applying \ref{axmc} constrained by \ref{sps} allows F to make the latter prediction under both preparations, thereby averting both fallacies.

\subsection{Effective classical bases for quantum agents}\label{qag}
For the purpose of the many puzzles related to the \frp\ Gedankenexperiment, it was quite adequate for us to consider agents who have well\hyp defined ``definite observational states'' or ``classical bases''. Here we will briefly speculate on how \ref{sps} could be extended to inform the reasoning of truly quantum agents, e.g.\ quantum computers. Let us begin by asking ourselves, ``How would $\wb$ have reasoned, if they were a quantum computer?''

While a human $\wb$ would be limited to reasoning based on the particular basis in which they have already observed $\lb$, a quantum computer would have the option to simply undo this measurement and use a more informative one (if the experimental design insisted on implementing the given measurement before starting to reason)! In this case, considering once again the state $\ket\phi_{\lb\rL}$ [Eq.~\eqref{eqphi}], re\hyp expressed as
\be
\sqrt{\fr16}\ket{\hb}_{\lb}\ket{\mathrm{ok}}_\rL+\sqrt{\fr56}\left(\sqrt{\fr15}\ket{\hb}_{\lb}+\sqrt{\fr45}\ket{\tb}_{\lb}\right)\ket{\mathrm{fail}}_\rL,
\ee
we see that the optimal way for $\wb$ to make error\hyp free predictions about $w$ is to perform an unambiguous state discrimination protocol between the states $\ket{\hb}_{\lb}$ and $\sqrt{\fr15}\ket{\hb}_{\lb}+\sqrt{\fr45}\ket{\tb}_{\lb}$. This would result in three mutually\hyp exclusive cases, leading respectively to the following inferences:
\begin{enumerate}
    \item $\lb$ is not in the state $\ket{\hb}_{\lb}$; therefore $w={}$fail.
    \item $\lb$ is not in the state $\sqrt{\fr15}\ket{\hb}_{\lb}+\sqrt{\fr45}\ket{\tb}_{\lb}$; therefore $w={}$ok.
    \item I am uncertain about the outcome of W's measurement.
\end{enumerate}
In this Gedankenexperiment, that would essentially be all $\wb$ would have to do. But imagine a case where this inference were just one step in a longer chain of reasoning. If basing subsequent inferences on the result of this one, we expect \ref{sps} to apply in the following sense: Operationally, the most general discrimination protocol of this kind can be implemented by interacting $\lb$ with some auxiliary system and then performing a sharp projective measurement on the latter, with respect to some basis. So long as the result of this inference is part of the reasoning premise, \ref{sps} could then be enforced by requiring that subsequent operations act incoherently relative to this basis.

In this sense, even though quantum computers are not restricted by fixed classical bases, we expect the soundness of their reasoning to depend on a variant of \ref{sps} applied on the ``effective classical bases'' determined by all the premises on which any given inference is based. We expect there to be a rich landscape of possible branching structures (including nested branches, opening and closing of the scope of a subordinate premise within that of another, etc.)\ and formal rules for the mutual compatibility of premises. We leave this avenue for future work.

\subsection{Arbitrating FR's casino dispute}\label{secop2}
To close this section on a lighter note, we revisit a puzzle \frp\ posed after proving their main theorem. Suppose their Gedankenexperiment is a casino's gambling game, with W the gambler and the rest of the agents casino employees. The rules stipulate a reward to the gambler if $r={}$heads, and a penalty if $r={}$tails.

On an instance where $\overline w=\overline{\mathrm{ok}}$ and $w={}$ok are observed, $\wb$ might reason, via the non\hyp\ref{sps} reasoning we reviewed in Section~\ref{frarg}, that $r$ must have landed tails. On the other hand, W might aver that their outcome $w={}$ok proves that the state of S before F's measurement was not $\ket\rightarrow_\rS$, and that therefore $r$ must have been heads. How is the dispute to be resolved?

By now, it is eminently clear that the game formulated as such is not operationally sound: its rules are based on extra\hyp operational details of interpretation. As such, both the parties are to blame for entering into an ambiguous arrangement. Yet, if we had to find the closest approximation to resolving the dispute, we would recommend applying the principle of virtual disclosure. As we saw in Section~\ref{secvd}, this would uphold $\fb$'s situational reasoning and reject that of $\wb$. In connection with the dispute, the court should then award the case to W, whose reasoning is compatible with the former.

Complications would arise, though, if $\fb$\m themselves a casino employee\m weighed in on the matter. $\fb$ might protest that they distinctly recall seeing the outcome $r={}$tails. Nevertheless, as we saw in Section~\ref{secsol}, in order for the game to work as designed, $\fb$ would have had to cede their cognitive agency entirely for its duration. The court should deem $\fb$ to thereby have relinquished their eligibility to act as a sound\hyp minded witness to any events that transpired at the time. Hence, overall our principles of \ref{sps} and virtual disclosure would lean in favour of the hapless gambler W. We might rule that the casino pay W their legal expenses (but not the stipulated reward); serve the casino a stern warning that it will be heavily penalized if it continues to lure unsuspecting gamblers into operationally ill\hyp defined games; and strongly advise W to gamble more responsibly henceforth.

\section{Summary and outlook}
When rational agents try to reason in situations where their very minds are in quantum superposition states, they can quickly run into paradoxes. If the ``Wigner's friend'' Gedankenexperiment offered a glimpse of this predicament, Frauchiger and Renner (FR)'s ``Wigners' friends'' extension \cite{FR18} has exposed its convoluted innards.

\frp\ interpreted their technical result as an impossibility for quantum mechanics to consistently describe its own use in reasoning. This definitive verdict, however, has been challenged by later authors. Unfortunately, much of the discourse on this topic has used the terminology and technical machinery of particular interpretations of quantum mechanics, such as the de Broglie\n Bohm interpretation, QBism, and consistent histories. Elsewhere, authors have brought to bear tools from the field of formal languages and modal logic.

We offer this work as a humble attempt to strip the issue of all interpretational elements, frame it operationally, study it in a language and formalism accessible to experts in the broad field of quantum information science, and make some novel contributions.

Our main contribution is an operational principle to aid rational agents in making inferences. Called \emph{superpositional solipsism} (\ref{sps}), this principle prescribes that an agent reason based on the premise that their current mental or observational state is definite and not part of a superposition.

We showed that \ref{sps} leads to operationally\hyp sound inferences in all instances where an agent has any agency to reason. Furthermore, we showed that a failure to respect \ref{sps} leads to a weakening in an agent's predictive power or, worse, a breakdown of rational soundness. We used the \frp\ Gedankenexperiment as a test case to uncover various attractive features of \ref{sps}, including its operational simplicity. We also used this opportunity to explore subtleties that might have been missed in past work, particularly those concerning rational agency.

We also demonstrated that neglecting \ref{sps} leads agents to reason based on fallacious assumptions that amount to a violation of the no\hyp signalling principle or of the operational informational completeness of the wavefunction. Moving beyond human\hyp like agents that have well\hyp defined ``observational states'', we offered some clues as to how to generalize this concept to the case of fully quantum agents such as quantum computers. We surmise that each situation where such an agent needs to make an inference naturally offers certain ``effective observational states'', associated with the situational premises. In future work, we hope to undertake the project of further formalizing the notions we introduced, particularly in connection with fully quantum agents. It might also be worth exploring if our concept of ``transmissive stability'' of logical statements can itself be elevated to some sort of axiomatic principle to replace \ref{sps}.

Also worth exploring are possible forms in which our ideas have already been presented by others before. We expect the framework of compatible observables and contextuality to be a likely source of such precedent: \ref{sps} could be equivalent to the principle that mutually\hyp incompatible observables cannot be simultaneously assigned definite values. Likewise, the notions of ``trust'' and ``scope'' introduced in \cite{NdR18} could be related to the mechanism through which \ref{sps} qualifies \frp's Axiom~\ref{axmc}.

Familiarity with the formalism might have trained physicists to expect coherent superpositional elements to be critical to quantum\hyp mechanical reasoning. Indeed, much of the response to \frp's inconsistency result, in scholarly literature and on internet discussions alike, has hinted that their fictional agents' situational negligence of alternate branches of the wavefunction might be the fatal fallacy underlying the inconsistency. Remarkably, our work shows that the fallacy lies at the opposite extreme: agents really \emph{should} be reasoning in staunch oblivion of branches of the wavefunction they don't experience\m \emph{even ones they expect to be present by design}!

Thus, the lay human intuition that reasons ``classically'' might well offer pragmatic and operational benefits. It would be interesting to investigate whether such benefits played a role in the evolution of our oft\hyp lamented classical intuition. More generally, it is intriguing why there are particular observational states, or ``classical bases'', with respect to which we experience consciousness. Could there be other, fundamentally different, ways in which conscious agents could evolve to experience the same reality? Would we notice these agents? What are the possible mechanisms through which an entity could acquire\m as we seem to have\m a ``working understanding'' of quantum mechanics despite lacking an ``intuitive understanding''? Inquiry into such questions might benefit from an approach similar to M\"uller's~\cite{Mueller20}.

A related avenue for foundational research, which has already received some attention, is regarding agency. In a world governed by Schr\"odinger's unitary dynamics, what does agency even mean, and how does it emerge and persist? If ``superagents'' were pulling our strings, would we be able to tell\m say, by virtue of ``miracle''\hyp like aberrations \`a la ``The Matrix''?

As a final note, we raise the question, ``Why should we expect quantum mechanics to have built\hyp in mechanisms to keep our reasoning sound?'' Perhaps it takes constant work and vigilance to get our thinking straight, but perhaps that's alright. Perhaps quantum mechanics \emph{can} consistently describe its own use\m if only we take care that it does!

\onecolumngrid

\appendix

\section{The FR Gedankenexperiment}
Here we will introduce the Gedankenexperiment devised by Frauchiger and Renner (\frp). We will paraphrase \frp's account in a way that, to the best of our judgement, does not conflict with their original intent.

The experiment involves four agents\m F, $\fb$, W, and $\wb$. Agents F and $\fb$ each have access to their respective quantum\hyp mechanical lab (whose contents will become clear by context). Meanwhile, W is a ``superagent'' over F in the following sense: W has full quantum\hyp mechanical control on the system L, defined as F's entire lab \emph{including its resident agent F}; likewise, $\wb$ is a superagent over $\fb$.

The experiment proceeds by repeating a set sequence of operations (``round'') indefinitely until a certain condition is met (like a ``while loop'' in a computer program). A typical round $n$ consists of the following operations, each carried out at a specific time index\footnote{Our time indices differ slightly from those of \frp, without loss of essence.} within the round:
\begin{enumerate}
\item At time $n$:00, Agent $\fb$ prepares a qubit R in the pure state
\be
\ket{\mathrm{init}}_\rR:=\sqrt{\fr13}\ket{\mathrm{heads}}_\rR+\sqrt{\fr23}\ket{\mathrm{tails}}_\rR,
\ee
where $\left\{\ket{\mathrm{heads}}_\rR,\ket{\mathrm{tails}}_\rR\right\}$ is an orthonormal basis over R's Hilbert space.
\item At time $n$:10, $\fb$ measures R in the $\left\{\ket{\mathrm{heads}}_\rR,\ket{\mathrm{tails}}_\rR\right\}$ basis using a device $\db$ (WLOG, a qubit identical to R), makes a mental note of the outcome, and proceeds to prepare another qubit, S, in a state conditional on the outcome. This results in the following multipartite state:
\be\label{eqpsi}
\ket{\psi}_{\rR\db\fb\rS}:=\sqrt{\fr13}\ket{\mathrm{heads}}_\rR\ket{\hb}_{\db}\ket{\hb}_{\fb}\ket{\downarrow}_\rS+\sqrt{\fr23}\ket{\mathrm{tails}}_\rR\ket{\tb}_{\db}\ket{\tb}_{\fb}\ket{\rightarrow}_\rS.
\ee
Here, $\left\{\ket{\uparrow}_\rS,\ket{\downarrow}_\rS\right\}$ is an orthonormal basis over S's Hilbert space, and $\ket{\rightarrow}_\rS=\sqrt{\fr12}\ket{\uparrow}_\rS+\sqrt{\fr12}\ket{\downarrow}_\rS$. We follow \frp\ in assigning a pure state\footnote{While this assignment probably does not reflect the complex workings of agents such as ourselves, it is benign and perfectly acceptable in the spirit of the Gedankenexperiment.} to $\fb$ at their having noted one or the other outcome; these two states of $\fb$ are also assumed to be fully informed about the outcome, and consequently, mutually orthogonal.
\item At time $n$:15, $\fb$ sends S to F's lab.
\item At time $n$:20, F uses a device D to measure S in the $\left\{\ket{\uparrow}_\rS,\ket{\downarrow}_\rS\right\}$ basis and makes a mental note of the outcome, resulting in the state\footnote{$\left\{\ket{\tfrac12}_\rD,\ket{-\tfrac12}_\rD\right\}$ and $\left\{\ket{\tfrac12}_\rF,\ket{-\tfrac12}_\rF\right\}$ are orthonormal bases over the Hilbert spaces of D and F, respectively.}
\begin{align}
\ket{\phi}_{\rR\db\fb\rS\rD\rF}:=&\sqrt{\fr13}\left(\ket{\mathrm{heads}}_\rR\ket{\hb}_{\db}\ket{\hb}_{\fb}\ket{\downarrow}_\rS\ket{-\tfrac12}_\rD\ket{-\tfrac12}_\rF\right.\nonumber\\
&\qquad\left.+\ket{\mathrm{tails}}_\rR\ket{\tb}_{\db}\ket{\tb}_{\fb}\ket{\downarrow}_\rS\ket{-\tfrac12}_\rD\ket{-\tfrac12}_\rF+\ket{\mathrm{tails}}_\rR\ket{\tb}_{\db}\ket{\tb}_{\fb}\ket{\uparrow}_\rS\ket{\tfrac12}_\rD\ket{\tfrac12}_\rF\right)\nonumber\\
=&:\sqrt{\fr13}\left(\ket{\hb}_{\lb}\ket{-\tfrac12}_\rL+\ket{\tb}_{\lb}\ket{-\tfrac12}_\rL+\ket{\tb}_{\lb}\ket{\tfrac12}_\rL\right),\label{eqphi}
\end{align}
where L consists of SDF and $\lb$ of R$\db\fb$, respectively. Note that $\bra{\hb}\left.\tb\right\rangle_{\lb}=\langle\tfrac12\ket{-\tfrac12}_\rL=0$.
\item At time $n$:30, $\wb$ uses device $\eb$ to measure $\lb$ with respect to the orthonormal basis $\left\{\ket{\overline{\mathrm{ok}}}_{\lb},\ket{\overline{\mathrm{fail}}}_{\lb}\right\}$, where $\ket{\overline{\mathrm{ok}}}_{\lb}:=\sqrt{\fr12}\left(\ket{\hb}_{\lb}-\ket{\tb}_{\lb}\right)$ and $\ket{\overline{\mathrm{fail}}}_{\lb}:=\sqrt{\fr12}\left(\ket{\hb}_{\lb}+\ket{\tb}_{\lb}\right)$. Subsequently, $\wb$ makes a mental note and an announcement of the outcome of this measurement. At this stage the state is
\be\label{eqxi}
\ket{\xi}_{\lb\eb\wb\rL}:=\sqrt{\fr23}\left[\ket{\overline{\mathrm{fail}}}_{\lb}\ket{\overline{\mathrm{fail}}}_{\eb}\ket{\overline{\mathrm{fail}}}_{\wb}\ket{-\tfrac12}_\rL+\fr12\left(\ket{\overline{\mathrm{fail}}}_{\lb}\ket{\overline{\mathrm{fail}}}_{\eb}\ket{\overline{\mathrm{fail}}}_{\wb}-\ket{\overline{\mathrm{ok}}}_{\lb}\ket{\overline{\mathrm{ok}}}_{\eb}\ket{\overline{\mathrm{ok}}}_{\wb}\right)\ket{\tfrac12}_\rL\right],
\ee
where the Hilbert spaces of $\eb$ and $\wb$ are spanned by orthonormal vectors corresponding to the eponymous ones of $\lb$.
\item At time $n$:40, W uses device E to measure L with respect to the orthonormal basis $\left\{\ket{\mathrm{ok}}_\rL,\ket{\mathrm{fail}}_\rL\right\}$, where $\ket{\mathrm{ok}}_\rL:=\sqrt{\fr12}\left(\ket{-\tfrac12}_\rL-\ket{\tfrac12}_\rL\right)$ and $\ket{\mathrm{fail}}_\rL:=\sqrt{\fr12}\left(\ket{-\tfrac12}_\rL+\ket{\tfrac12}_\rL\right)$. Finally, W makes a mental note and an announcement of the outcome. The final state is
\begin{align}
\ket{\zeta}_{\lb\eb\wb\rL\rE\rW}:=&\sqrt{\fr13}\left[\vphantom{\sqrt{\fr12}}\ket{\overline{\mathrm{fail}}}_{\lb}\ket{\overline{\mathrm{fail}}}_{\eb}\ket{\overline{\mathrm{fail}}}_{\wb}\left(\ket{\mathrm{ok}}_\rL\ket{\mathrm{ok}}_\rE\ket{\mathrm{ok}}_\rW+\ket{\mathrm{fail}}_\rL\ket{\mathrm{fail}}_\rE\ket{\mathrm{fail}}_\rW\right)\right.\nonumber\\
&\qquad+\left.\fr12\left(\ket{\overline{\mathrm{ok}}}_{\lb}\ket{\overline{\mathrm{ok}}}_{\eb}\ket{\overline{\mathrm{ok}}}_{\wb}-\ket{\overline{\mathrm{fail}}}_{\lb}\ket{\overline{\mathrm{fail}}}_{\eb}\ket{\overline{\mathrm{fail}}}_{\wb}\right)\left(\ket{\mathrm{ok}}_\rL\ket{\mathrm{ok}}_\rE\ket{\mathrm{ok}}_\rW-\ket{\mathrm{fail}}_\rL\ket{\mathrm{fail}}_\rE\ket{\mathrm{fail}}_\rW\right)\right]\nonumber\\
=&:\sqrt{\fr1{12}}\left[\ket{\overline{\mathbf{ok}}}_{\lb\eb\wb}\ket{\mathbf{ok}}_{\rL\rE\rW}-\ket{\overline{\mathbf{ok}}}_{\lb\eb\wb}\ket{\mathbf{fail}}_{\rL\rE\rW}+\ket{\overline{\mathbf{fail}}}_{\lb\eb\wb}\ket{\mathbf{ok}}_{\rL\rE\rW}+3\ket{\overline{\mathbf{fail}}}_{\lb\eb\wb}\ket{\mathbf{fail}}_{\rL\rE\rW}\right],
\end{align}
where all the newly\hyp named vectors have contextually\hyp evident meanings.
\end{enumerate}
The outcomes of the measurements carried out by $\fb$, F, $\wb$, and W are represented in \cite{FR18} by the symbols $r$, $z$, $\overline w$ and $w$. The above sequence of operations is repeated until both W and $\wb$ observe their respective ``ok'' outcomes in the same round. As can be seen from the form of the final state $\ket{\zeta}_{\lb\eb\wb\rL\rE\rW}$, this condition has the nonzero probability $\fr1{12}$ of occurring in any given round. It is therefore certain to be satisfied on some finite round $n$, whereat the experiment halts.


\end{document}